\documentclass[showpacs,superscriptaddress,nofootinbib,amssymb,amsmath,twocolumn]{revtex4}
\usepackage{txfonts}
\usepackage[dvips]{graphicx}
\usepackage{color,latexsym,epsfig,bm,times,subfigure}
\usepackage[ps2pdf,bookmarks=true,colorlinks,linkcolor=blue,urlcolor=green,citecolor=red]{hyperref}

\renewcommand\d{\delta}

\renewcommand\r{\rho}

\renewcommand\t{\tau}

\renewcommand\c{\chi}
\renewcommand\j{\psi}

\newcommand\e{\epsilon}
\newcommand\g{\gamma}

\newcommand\m{\mu}

\newcommand\p{\pi}

\newcommand\s{\sigma}
\newcommand\f{\phi}
\newcommand\w{\eta}
\newcommand\D{\Delta}

\newcommand{\fig}[1]{Fig.~\ref{#1}}
\newcommand{\eq}[1]{Eq.~(\ref{#1})}
\newcommand{\eqs}[2]{Eqs.~(\ref{#1})-(\ref{#2})}

\newcommand\lb{\left(}
\newcommand\rb{\right)}
\newcommand\ls{\left[}
\newcommand\rs{\right]}

\newcommand{\lan}{\langle}
\newcommand{\ran}{\rangle}

\newcommand{\non}{\nonumber\\}

\newcommand{\br}{{\mathbf r}}

\newcommand{\bv}{{\bf v}}

\newcommand{\bB}{{\bf B}}
\newcommand{\bE}{{\bf E}}

\newcommand{\rp}{{\rm RP}}

\renewcommand{\part}{{\rm part}}
\begin{document}

\title{Electric Fields and Chiral Magnetic Effect in Cu + Au Collisions}
\author{Wei-Tian Deng}
\affiliation{School of physics, Huazhong University of Science and Technology, Wuhan 430074, China}
\author{Xu-Guang Huang}
%\email{huangxuguang@fudan.edu.cn}
\affiliation{Physics Department and Center for Particle Physics and Field Theory, Fudan University, Shanghai 200433, China.}

\date{\today}

\begin{abstract}
The non-central Cu + Au collisions can create strong out-of-plane magnetic fields and in-plane electric fields. By using the HIJING model, we study the general properties of the electromagnetic fields in Cu + Au collisions at $200$ GeV and their impacts on the charge-dependent two-particle correlator $\gamma_{q_1q_2}=\lan\cos(\phi_1+\phi_2-2\psi_\rp)\ran$ (see main text for definition) which was used for the detection of the chiral magnetic effect (CME). Compared with Au + Au collisions, we find that the in-plane electric fields in Cu + Au collisions can strongly suppress the two-particle correlator or even reverse its sign if the lifetime of the electric fields is long. Combining with the expectation that if $\gamma_{q_1q_2}$ is induced by elliptic-flow driven effects we would not see such strong suppression or reversion, our results suggest to use Cu + Au collisions to test CME and understand the mechanisms that underlie $\g_{q_1q_2}$.
\end{abstract}


\maketitle
\section{Introduction}
Over the past few years, there has been increasing interest in the quantum-anomaly-related transport phenomena and their experimental signals in heavy-ion collisions. Such anomalous transports include chiral magnetic effect (CME)~\cite{Kharzeev:2007jp,Fukushima:2008xe,Vilenkin:1980fu}, chiral vortical effect (CVE)~\cite{Vilenkin:1979ui,Kharzeev:2007tn,Son:2009tf}, chiral separation effect (CSE)~\cite{Son:2004tq,Metlitski:2005pr}, chiral electric separation effect (CESE)~\cite{Huang:2013iia,Jiang:2014ura,Pu:2014cwa,Pu:2014fva}, etc. All these effects involve a net chiral imbalance in the quark-gluon plasma (QGP) and their occurrence, if confirmed, provides us hitherto unique evidence for the local P and CP violation in QGP. For review on the anomalous transports in heavy-ion collisions, see for example Refs~\cite{Bzdak:2012ia,Kharzeev:2013ffa,Liao:2014ava}.

The CME in QGP can be neatly expressed as
\begin{eqnarray}
\label{cme}
{\bf J}=\s_5\bB\;\;{\rm with}\;\;\s_5=N_c\sum_f\frac{ q_f^2 \m_5}{2\p^2},
\end{eqnarray}
where the summation is over all light quark flavors, $N_c$ is the number of colors, $\m_5$ is the chiral chemical potential, $q_f$ is the electric charge of quark flavor $f$, and $\bB$ is a magnetic field. Equation (\ref{cme}) represents an electric current induced by a magnetic field in QGP with net chirality (characterized by $\m_5$). In real heavy-ion collisions, the magnetic fields are generated mostly in the direction perpendicular to the reaction plane (For discussions about the event-by-event fluctuation of the magnetic field orientation, see Refs.~\cite{Bloczynski:2012en}), therefore the CME is expected to induce a charge separation with respect to the reaction plane which may be detected, as proposed by Voloshin~\cite{Voloshin:2004vk}, via the two-particle correlator
\begin{eqnarray}
\label{gam}
\g_{q_1q_2}=\lan\cos(\f_1+\f_2-2\j_\rp)\ran,
\end{eqnarray}
where $q_1$ and $q_2$ are charges of particles $1$ and $2$, $\f_1, \f_2$, and $\j_\rp$ are the azimuthal angles for the particles $1$, $2$, and the reaction plane, respectively, and the average in \eq{gam} is taken over events. Negative same-sign (SS) correlator $\g_{\rm SS}\equiv(\g_{++}+\g_{--})/2$ and positive opposite-sign (OS) correlator $\g_{\rm OS}\equiv\g_{+-}$ could constitute a strong evidence for the occurrence of CME. The measurements have been carried out by STAR Collaboration~\cite{Abelev:2009ac,Abelev:2009ad,Wang:2012qs,Adamczyk:2014mzf} at RHIC for Au + Au and Cu + Cu collisions and ALICE Collaboration~\cite{Abelev:2012pa} at LHC for Pb + Pb collisions. The experimental results of $\g_{q_1q_2}$ showed consistent behavior with the expectation of CME. However, there still remain debates~\cite{Wang:2009kd,Bzdak:2009fc,Bzdak:2010fd,Liao:2010nv,Pratt:2010zn,Schlichting:2010qia} on the CME interpretation of the data because $\g_{q_1q_2}$ may have contributions from other effects, potentially the elliptic-flow ($v_2$) driven ones, e.g., the transverse momentum conservation (TMC)~\cite{Pratt:2010zn,Bzdak:2010fd} and local charge conservation (LCC)~\cite{Schlichting:2010qia}. (If one measures the difference between the SS and OS correlators, the TMC contribution can be subtracted as it is charge independent, but the LCC remains to contribute.) It was proposed~\cite{Voloshin:2010ut} to use the central U + U collisions which are expected to have sizable elliptic flows but no magnetic fields to disentangle the CME and $v_2$ driven contributions. Preliminary results for $\g_{q_1q_2}$ in U + U collisions have been reported by STAR collaboration~\cite{Wang:2012qs}, but there are still uncertainties, see discussions in Ref.~\cite{Bloczynski:2013mca}.

%%%%%%%%%%%%%%%%%%%%%%%%%%%%%%%%%%%%%%%%%%%%%%%%%%%%%%%%%%%%%%%%%%%%%%%
\begin{figure}[!htb]
\begin{center}
\includegraphics[width=8cm]{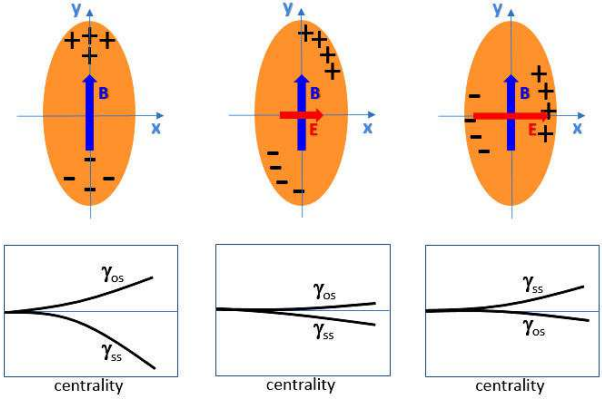}
\caption{(Color online) Illustration of the charge dipole induced by CME and electric field in Cu + Au collisions. (Left): In-plane electric field is negligible. The dipole is driven by CME, and thus is along the magnetic field direction. (Middle): Weak electric field. The in-plane electric field drives the dipole to skew from the magnetic field direction, and thus reduce the magnitudes of both $\g_{\rm OS}$ and $\g_{\rm SS}$. (Right): Strong electric field. The in-plane electric field drives a large in-plane component of the charge dipole, and thus the signs of $\g_{\rm OS}$ and $\g_{\rm SS}$ will be reversed.}
\label{illu}
\end{center}
\end{figure}
%%%%%%%%%%%%%%%%%%%%%%%%%%%%%%%%%%%%%%%%%%%%%%%%%%%%%%%%%%%%%%%%%%%%%%%
The purpose of this paper is to propose to use another colliding system, that is, the Cu + Au collisions at RHIC to test CME-induced and the $v_2$-driven contributions to $\g_{q_1q_2}$. The idea is the following. If $\g_{q_1q_2}$ is dominated by $v_2$-driven effects, we expect that $\g_{q_1q_2}$ (more precisely, $\D\g\equiv\g_{\rm OS}-\g_{\rm SS}$, because the directed flow in Cu + Au collisions may contribute to both $\g_{\rm OS}$ and $\g_{\rm SS}$. This contribution does not exist in Au + Au collisions.) would not change too much from Au + Au collisions to Cu + Au collisions. (A plausible guess would be that $\D\g$ as a function of centrality in Cu + Au collisions lie between that in Cu + Cu and Au + Au collisions). However, on the other hand, if $\g_{q_1q_2}$
is driven by CME, then the correlator in Cu + Au collisions will be very different from that in Au + Au collisions. This is because that, in noncentral Cu + Au collisions, due to the asymmetric collision geometry, there can generate electric fields pointing from Au to Cu nuclei which can induce an in-plane charge separation in addition to the out-of-plane charge separation due to CME. Thus the resulting charge dipolar distribution will deviate from the out-of-plane direction, and as a consequence, both the SS and OS correlators will descend, see \fig{illu} for an illustration. If the in-plane electric fields are so strong (or somehow equivalently, if their lifetime is long. We will not distinguish these two situations.) that the resulting charge dipole is mainly along the in-plane direction, one should observe a positive $\g_{\rm SS}$ and a negative $\g_{\rm OS}$ as opposite to that in Au + Au collisions, see the most right panel in \fig{illu}. Thus if the measurement in Cu + Au collisions shows a strong reduction or even sign-reversion for $\g_{\rm OS}$ and $\g_{\rm SS}$ comparing to that in Au + Au collisions, it will be a hint for the EM-field (i.e., CME plus electric field) interpretation of $\g_{q_1q_2}$. Otherwise, it will indicate a $v_2$-driven interpretation of $\g_{q_1q_2}$.

The paper is organized as following. In Sec.~\ref{emfield} we will report our numerical simulations for some general properties of the EM fields in Cu + Au collisions. In Sec.~\ref{seccorr} we will then use these results to study the correlator $\g_{q_1q_2}$ in Cu + Au collisions. Finally we summarize and discuss in Sec.~\ref{concl}. Throughout this paper, we use the natural units $\hbar=k_B=c=1$.

\section{Electromagnetic fields in Cu + Au collisions}\label{emfield}
\subsection{Setups}
The aim of this section is to give an event-by-event calculation of the electromagnetic fields in Cu + Au collisions. The EM fields in symmetric collisions like Au + Au at RHIC and Pb + Pb at LHC have been investigated in Refs.~\cite{Skokov:2009qp,Voronyuk:2011jd,Bzdak:2011yy,Deng:2012pc}.
We focus on the fields at the initial time $t=0$ which we define as the moment when the two colliding nuclei are maximally overlapped.
We use the Li\'enard-Wiechert potentials to calculate the fields:
\begin{eqnarray}
\label{LWE}
e\bE(t,\br)&=&\frac{e^2}{4\p}\sum_n Z_n({\bf R}_n)\frac{{\bf R}_n-R_n\bv_n}{(R_n-{\bf R}_n\cdot\bv_n)^3}(1-v_n^2),\\
\label{LWB}
e\bB(t,\br)&=&\frac{e^2}{4\p}\sum_n Z_n({\bf R}_n)\frac{\bv_n\times{\bf R}_n}{(R_n-{\bf R}_n\cdot\bv_n)^3}(1-v_n^2),
\end{eqnarray}
where $e>0$ is the proton charge, ${\bf R}_n=\br-\br_n(t)$ is the relative
position of the field point $\br$ to the $n$th proton at time $t$, $\br_n(t)$, and $\bv_n$ is the velocity of the $n$th proton.
The summations run over all protons in the projectile (we set it as Cu nucleus) and target (Au nucleus). The Li\'enard-Wiechert potentials are singular at ${\bf R}_n={\bf 0}$. In order to regulate the singularities we treat protons as uniformly charged spheres with radius $R_p$. This is reflected in the charge number factor $Z_n(\bf{R}_n)$ in Eqs.~(\ref{LWE}) and (\ref{LWB}): when the field point locates outside the $n$th proton (in the rest frame of the proton) we set $Z_n=1$, otherwise we cut the proton into two spherical domes at ${\bf R}_n$ and sum the contributions of these two domes. The in-medium charge radius $R_p$ of proton is unknown, we use the recent measurement of the rms charge radius of proton in vacuum, $R_p=0.8775$ fm ~\cite{Mohr:2012tt}. Varying $R_p$ from $0.7$ fm to $0.9$ fm brings about $10\%$ shift to the numerical results but no qualitative conclusion is altered. We set the projectile beam direction as $z$ axis and the $x$-axis is along the impact parameter vector (pointing from Au nucleus to Cu nucleus) so that the reaction plane is the $x$-$z$ plane. Finally, the positions of nucleons in the rest frame of a nucleus are sampled according to the Woods-Saxon distribution.

In real collisions, not only the strengths of the EM fields, but also their orientations with respect to the matter geometry (which we refer to as the participant planes) vary from event to event. We define the $n$th participant plane angle $\j_n$ and eccentricity $\e_n$ as (see e.g. Refs.~\cite{Teaney:2010vd,Qiu:2011iv}):
\begin{eqnarray}
\label{part:def}
\e_1e^{i\j_1}&=&-\frac{\int d^2\br_\perp\r(\br_\perp) r_\perp^3 e^{i\f}}{{\int d^2\br_\perp\r(\br_\perp) r_\perp^3}},\non
\e_n e^{in\j_n}&=&-\frac{\int d^2\br_\perp\r(\br_\perp) r_\perp^n e^{in\f}}{\int d^2\br_\perp\r(\br_\perp) r_\perp^n}\;\;\;\; {\rm for}\;\; n>1,\nonumber
\end{eqnarray}
where $\r(\br_\perp)$ is the participant density projected onto the transverse plane. For each event, we use the HIJING model~\cite{Wang:1991hta,Deng:2010mv,Deng:2010xg} to describe the collision dynamics and determine the centrality, the density, the angles $\j_n$, etc, and apply \eqs{LWE}{LWB} to calculate the EM fields.

\subsection{Results}
\begin{figure}[!htb]
\begin{center}
\includegraphics[width=7cm]{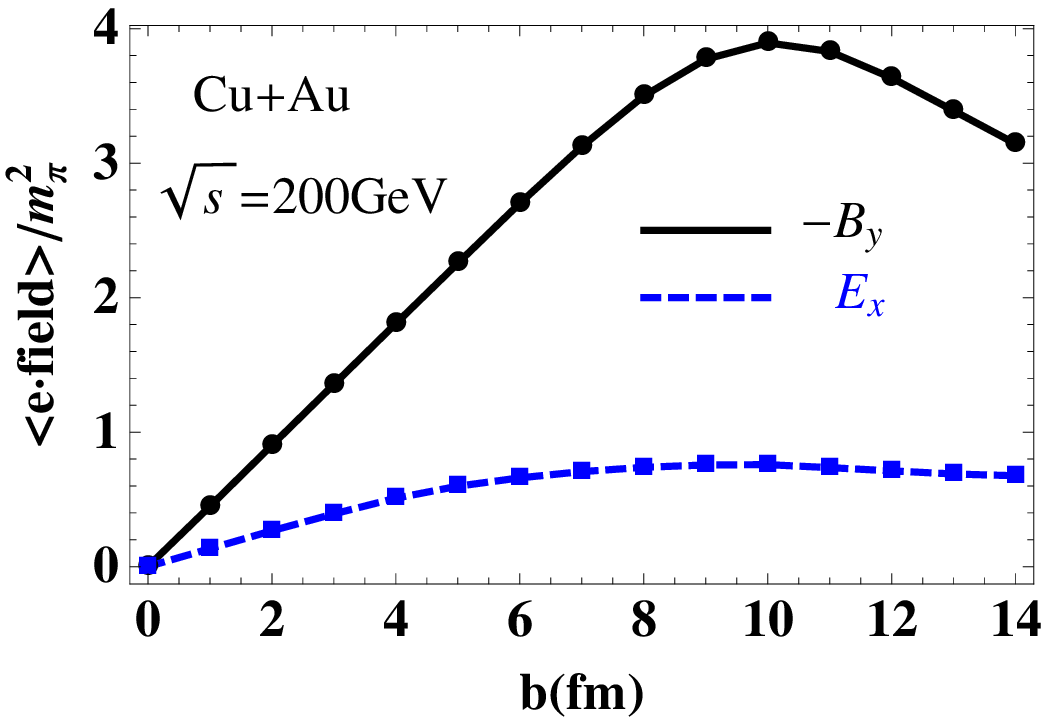}
\includegraphics[width=7cm]{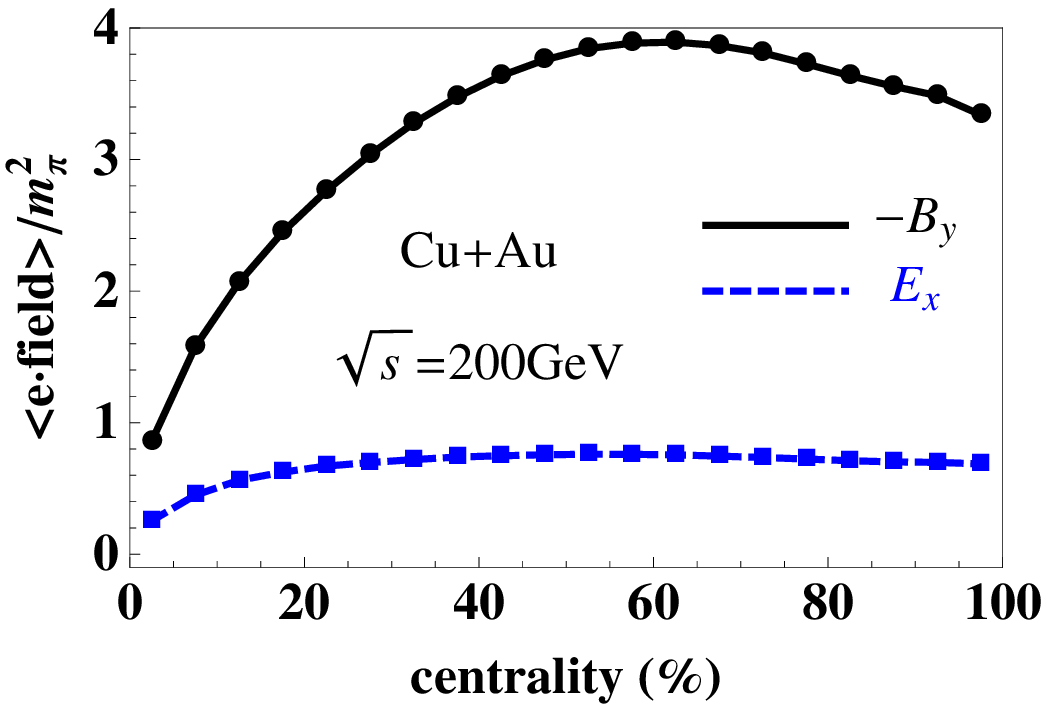}
\caption{(Color online) The event averaged magnetic and electric fields for Cu + Au collisions at the center of mass of the collision region as functions of impact parameter (upper panel) and centrality (lower panel).}
\label{average_c}
\end{center}
\end{figure}

\begin{figure}[!htb]
\begin{center}
\includegraphics[width=7cm]{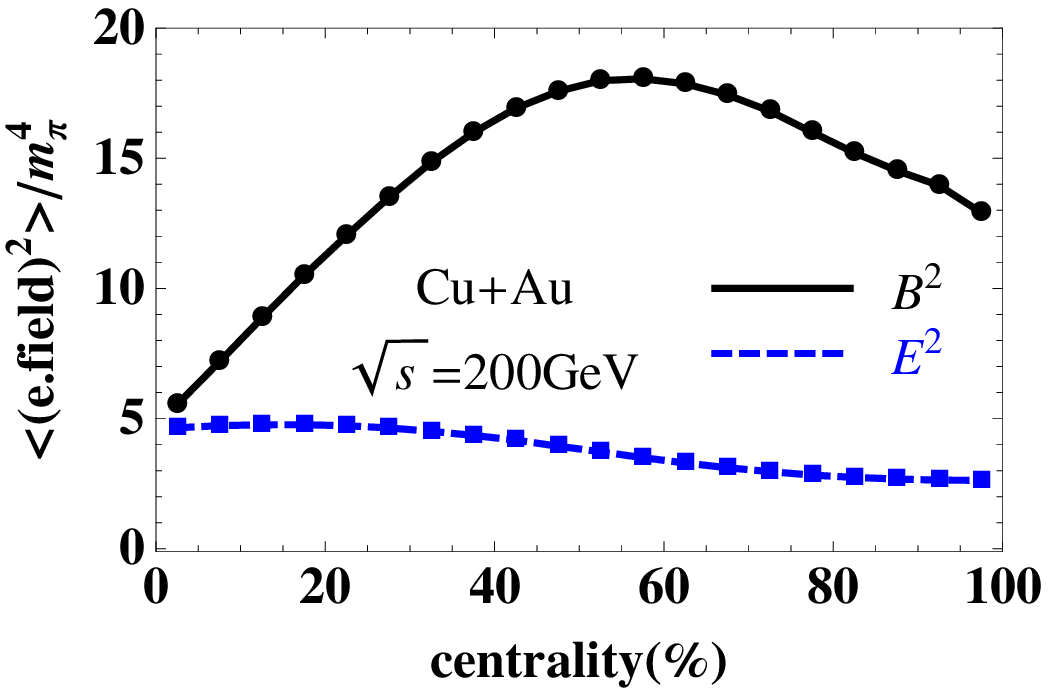}
\caption{(Color online) $\lan \bB^2\ran$ and $\lan \bE^2\ran$ in Cu + Au collisions as functions of centrality.}
\label{average_b2}
\end{center}
\end{figure}

{\it 1. Impact parameter and centrality dependence.---}
In \fig{average_c} we show the impact parameter and centrality dependence of the event averaged EM fields at the mass center of the collision region, where the center of mass is defined through $\br^{cm}_\perp=\int d^2\br_\perp\r(\br_\perp) \br_\perp$. First, the strength of the event averaged magnetic field (which is along the $-y$ direction) in Cu + Au collisions at 200 GeV is comparable to that in Au + Au collisions~\cite{Deng:2012pc}. Second, as a consequence of the asymmetry of the collision geometry, we find a strong event averaged electric field pointing from the Au nucleus to Cu nucleus. The strength of this electric field is smaller than the magnetic field, but still very large. The effect of this strong electric field on the two-particle correlator $\g_{q_1q_2}$ will be studied in Sec.~\ref{seccorr}. In \fig{average_b2}, we show $\lan \bB^2\ran$ and $\lan \bE^2\ran$ in Cu + Au collisions as functions of centrality. They reflect the event-by-event fluctuation of the strengths of the EM fields. This is particularly clear in central collisions where although the event-averaged EM fields vanish, the event-averaged squares of them are finite.
%%%%%%%%%%%%%%%%%%%%%%%%%%%%%%%%%%%%%%%%%%%%%%%%%%%%%%%%%%%%%%%%%%%%%%%
%%%%%%%%%%%%%%%%%%%%%%%%%%%%%%%%%%%%%%%%%%%%%%%%%%%%%%%%%%%%%%%%%%%%%%%
%%%%%%%%%%%%%%%%%%%%%%%%%%%%%%%%%%%%%%%%%%%%%%%%%%%%%%%%%%%%%%%%%%%%%%%
%%%%%%%%%%%%%%%%%%%%%%%%%%%%%%%%%%%%%%%%%%%%%%%%%%%%%%%%%%%%%%%%%%%%%%%

\begin{figure}[!htb]
\begin{center}
\includegraphics[width=7cm]{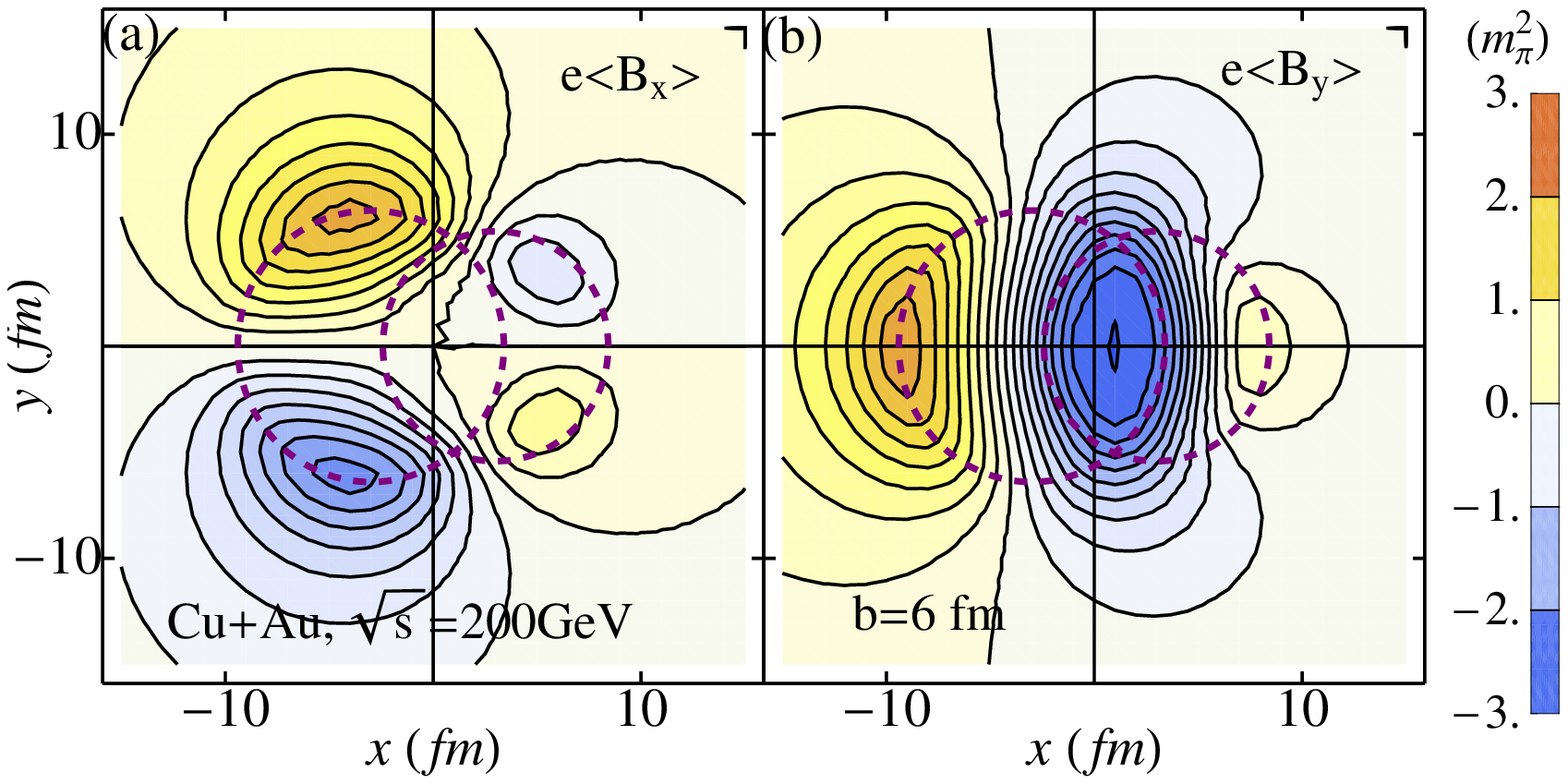}
\includegraphics[width=7cm]{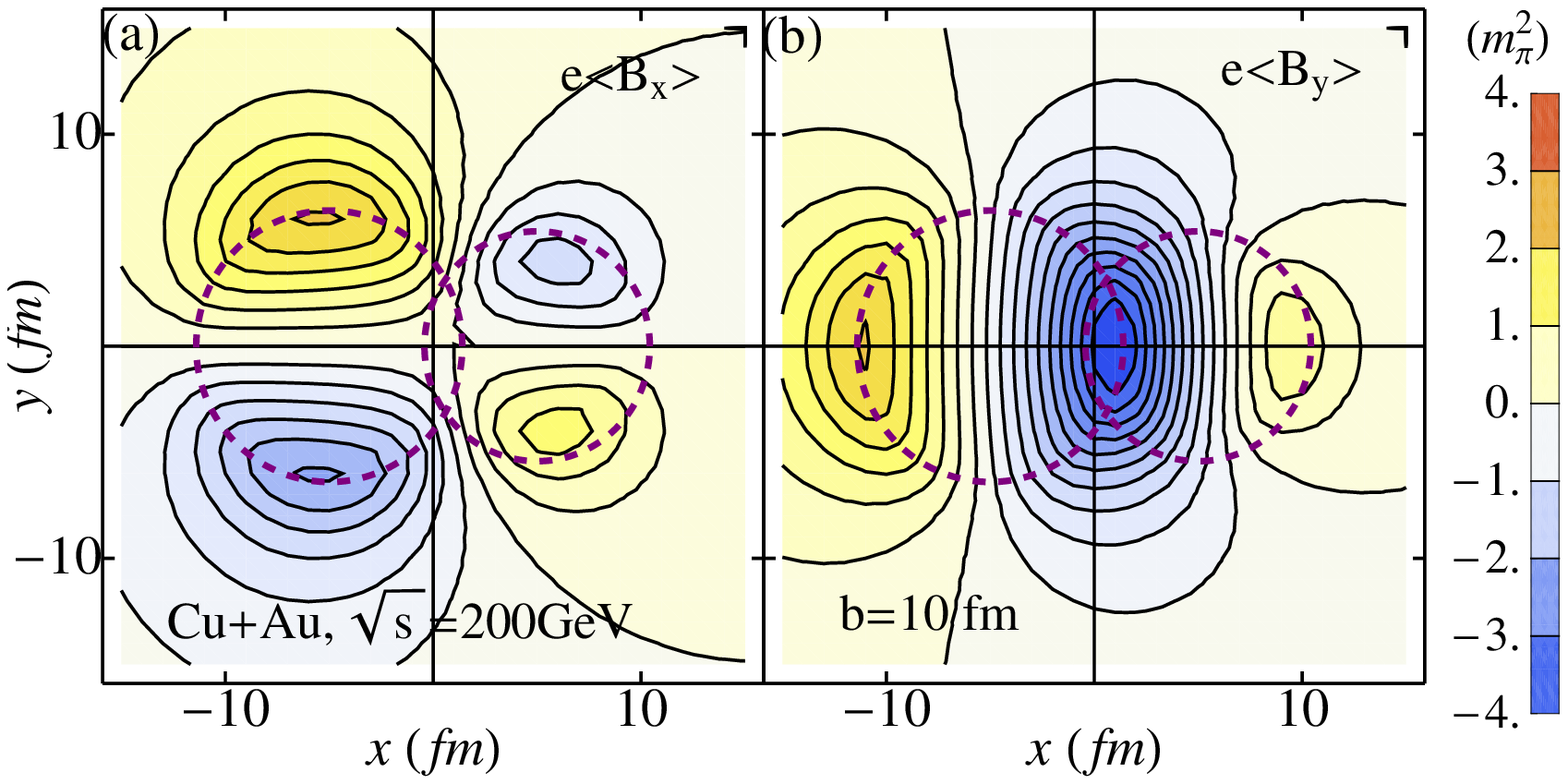}
\caption{(Color online) The spacial distribution of the event averaged magnetic field for Cu + Au collisions on the transverse plane. Upper panels: $b=6$ fm. Lower panels: $b=10$ fm.} \label{spaceb}
\end{center}
\end{figure}

\begin{figure}[!htb]
\begin{center}
\includegraphics[width=7cm]{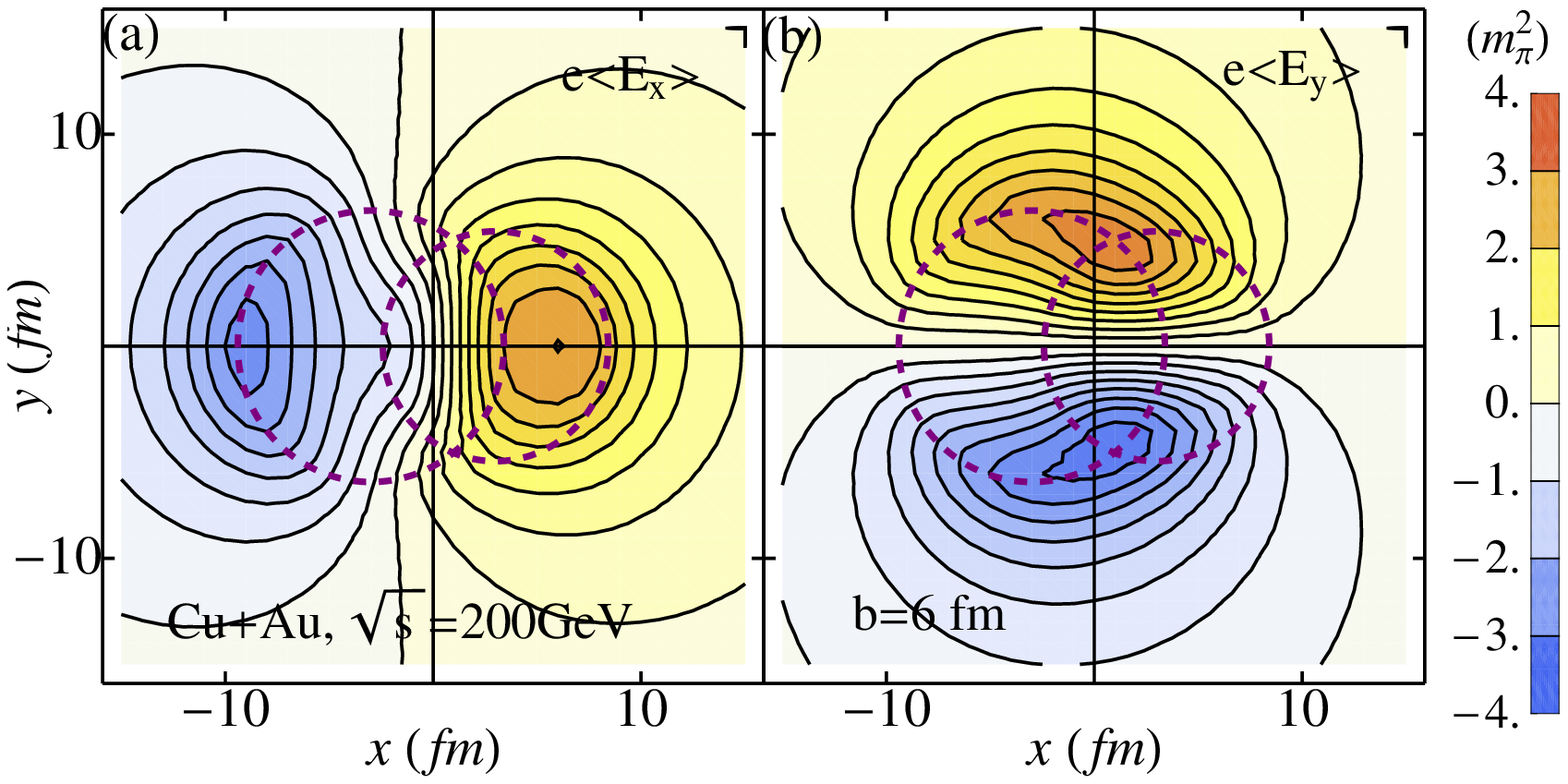}
\includegraphics[width=7cm]{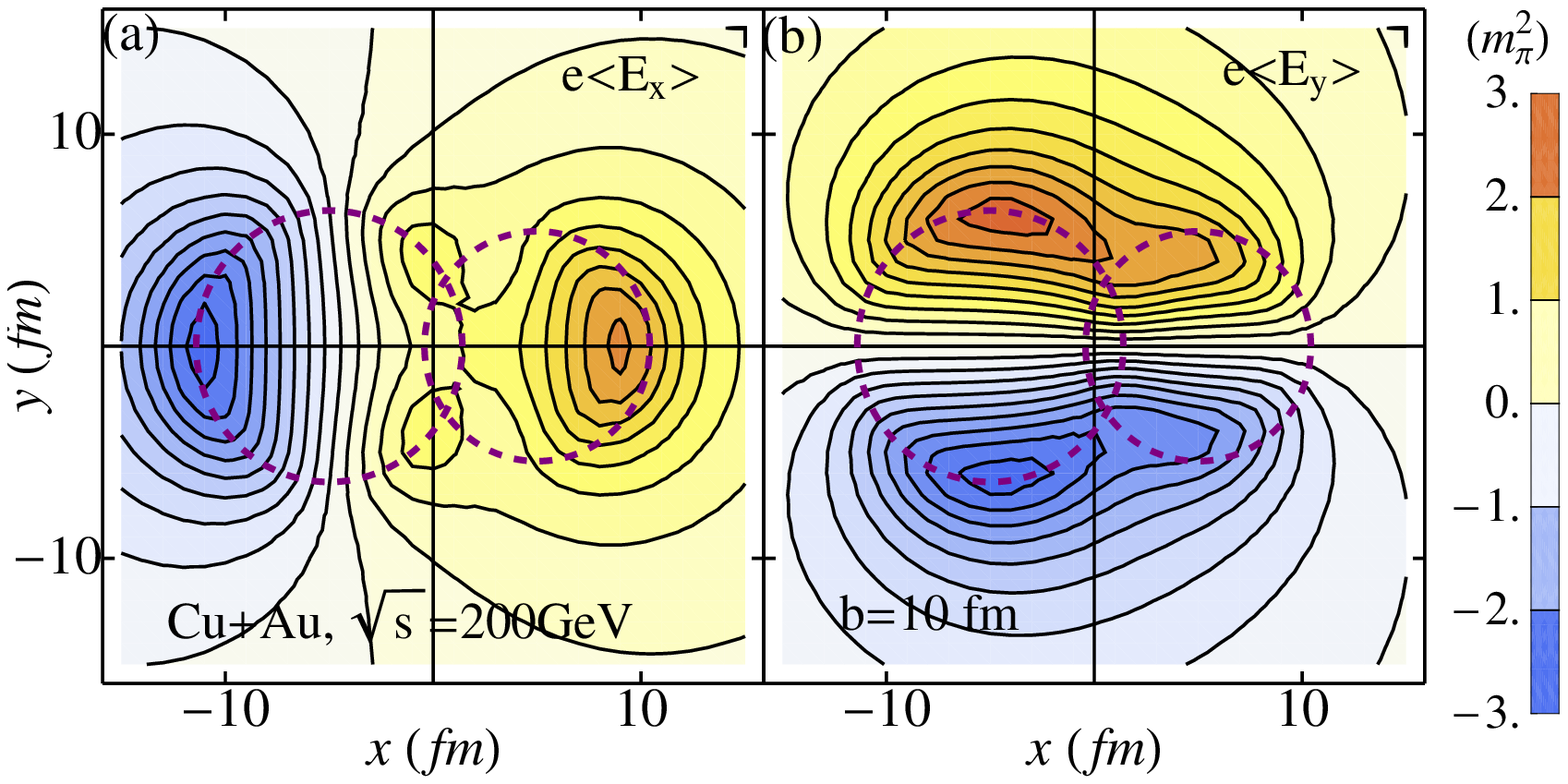}
\caption{(Color online) The spacial distribution of the event averaged electric field for Cu + Au collisions on the transverse plane. Upper panels: $b=6$ fm. Lower panels: $b=10$ fm.}
\label{spacee}
\end{center}
\end{figure}

{\it 2. Spacial distribution.---}
In \fig{spaceb} and \fig{spacee} we show the spacial distributions (on the transverse plane) of the event averaged magnetic and electric fields for Cu + Au collisions at 200 GeV. Obviously, the fields are very inhomogeneous in the overlapping region. Comparing to the Au + Au collisions~\cite{Deng:2012pc}, the asymmetric geometry of the Cu + Au collisions brings asymmetries to the spacial distributions of the fields. This is most evident for $\lan E_x\ran$ at $b=10$ fm: $\lan E_x\ran$ is positive everywhere in the overlapping region in Cu + Au collisions.
%%%%%%%%%%%%%%%%%%%%%%%%%%%%%%%%%%%%%%%%%%%%%%%%%%%%%%%%%%%%%%%%%%%%%%%
%%%%%%%%%%%%%%%%%%%%%%%%%%%%%%%%%%%%%%%%%%%%%%%%%%%%%%%%%%%%%%%%%%%%%%%
%%%%%%%%%%%%%%%%%%%%%%%%%%%%%%%%%%%%%%%%%%%%%%%%%%%%%%%%%%%%%%%%%%%%%%%
%%%%%%%%%%%%%%%%%%%%%%%%%%%%%%%%%%%%%%%%%%%%%%%%%%%%%%%%%%%%%%%%%%%%%%%

\begin{figure}[!htb]
\begin{center}
\includegraphics[width=7cm]{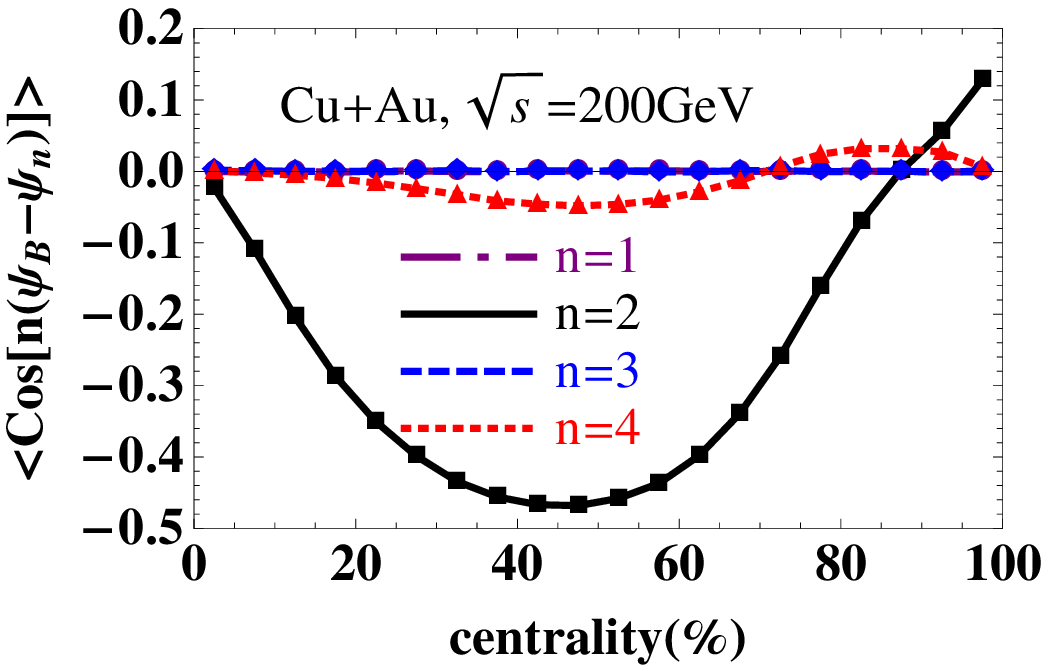}
\includegraphics[width=7cm]{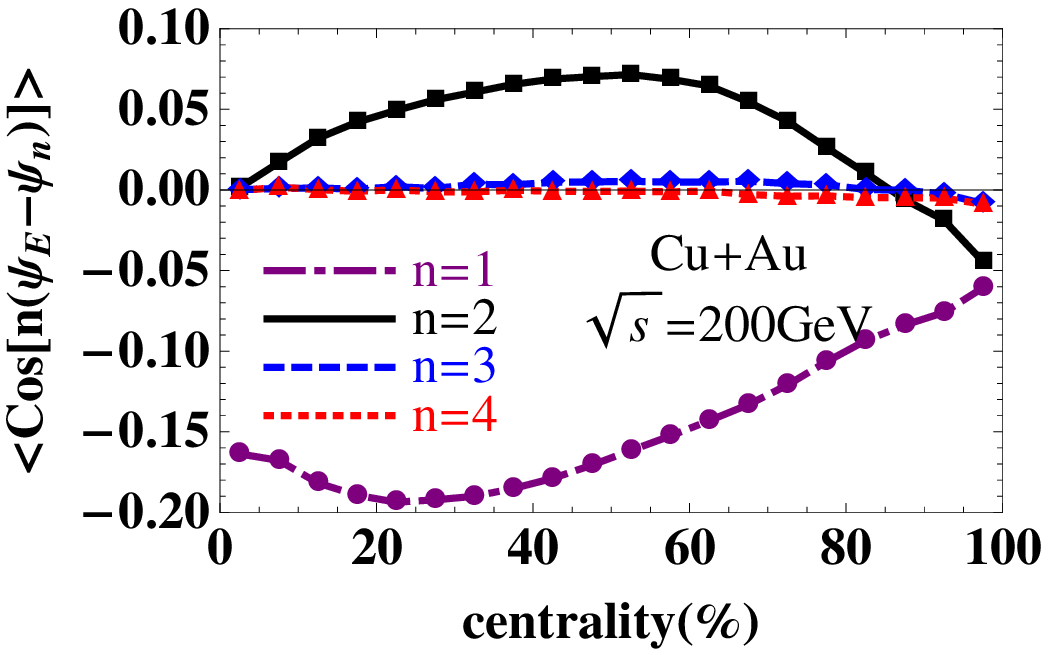}
\caption{(Color online) The azimuthal correlations $\lan\cos{[n(\j_B-\j_n)]}\ran$ and $\lan\cos{[n(\j_E-\j_n)]}\ran$ as functions of centrality.}
\label{azimu}
\end{center}
\end{figure}

{\it 3. Azimuthal correlations.---} In \fig{azimu} we show the numerical results for the correlations $\lan\cos{[n(\j_{E,B}-\j_n)]}\ran$ ($n=1-4$) which measure how strongly the azimuthal direction of the electric field (or magnetic field) correlate to the $n$th harmonic angle $\j_n$ of the participants. See Refs.~\cite{Bloczynski:2012en,Bloczynski:2013mca} and Sec.~\ref{seccorr} for the reasoning of studying these correlations. Similar with what has been observed in Au + Au collisions~\cite{Bloczynski:2012en}, the correlations between $\j_B$ and the odd harmonics, $\j_1,\j_3$, are practically zero, while the correlations between $\j_B$ and the even harmonics, $\j_2,\j_4$, are nonzero for moderate centralities but get suppressed compared with the case in the hard-sphere limit without event-by-event fluctuation (where, e.g., $\lan\cos{[2(\j_{B}-\j_2)]}\ran=-1$). For the electric field case, we observe that $\j_E$ is practically not correlated to $\j_3$ and $\j_4$, but there is a strong back-to-back correlation between $\j_E$ and $\j_1$ (the same as that in Au + Au collisions~\cite{Bloczynski:2012en}). Differently from the Au + Au case, in noncentral Cu + Au collisions there is a clear positive correlation between $\j_E$ and $\j_2$ signaling a persistent in-plane electric field. We note that at centrality $\gtrsim 90\%$, the correlations $\lan\cos{[2(\j_{B}-\j_2)]}\ran$ and $\lan\cos{[2(\j_{E}-\j_2)]}\ran$ change signs. This is because at that centrality region the most probably collisions involve only two participants distributed in the in-plane direction, and thus lead $\j_2$ to be mostly perpendicular to the reaction plane.
%%%%%%%%%%%%%%%%%%%%%%%%%%%%%%%%%%%%%%%%%%%%%%%%%%%%%%%%%%%%%%%%%%%%%%%
%%%%%%%%%%%%%%%%%%%%%%%%%%%%%%%%%%%%%%%%%%%%%%%%%%%%%%%%%%%%%%%%%%%%%%%

\section{Electric fields and two-particle correlators}\label{seccorr}
\subsection{Formalism}
Both the electric field and the magnetic field can drive charges to flow via the normal Ohm's law and the CME, respectively, and can build certain charge distribution patterns. In general, we can write down a single-particle distribution per rapidity as
\begin{eqnarray}
\label{single_dis}
f_1(q,\f)&=&\frac{N_q}{2\p}\ls1+2v_1^0\cos(\f-\j_1)\right.
\non&&\left.+2q d_E\cos(\f-\j_E)+2\c q d_B\cos(\f-\j_B)\right.\non&&\left.+{\rm higher\;\; harmonics}\rs,
\end{eqnarray}
where $q=\pm$ is the charge of the particle, $N_q$ is the number of hadrons with charge $q$ per rapidity, $v_1^0$ is the directed flow parameter at $\bE=\bB={\bf0}$, $d_E$ and $d_B$ characterize the strength of the dipole induced by $\bE$ and $\bB$, respectively, (We will estimate $d_{E}$ and $d_{B}$ latter), and $\c=\pm$ is a random variable accounting for the fact that in a given event there may be a sphaleron or an
anti-sphaleron transition resulting in charge separation parallel or anti-parallel to the $\bB$-field via CME. The event average of $\c$ is zero.
Equation (\ref{single_dis}) accounts for three kinds of ``flows": the directed flow along the azimuthal angle $\j_1$, the CME along $\j_B$, and the normal electric conduction along $\j_E$ .

From the single-particle distribution (\ref{single_dis}), we first observe that the directed flow of charge $q$ reads
\begin{eqnarray}
v_1(q)&=&v_1^0+q\lan d_E\cos(\j_E-\j_1)\ran+q\lan\c d_B\cos(\j_B-\j_1)\ran\non
&=&v_1^0+q\lan d_E\cos(\j_E-\j_1)\ran,
\end{eqnarray}
where $\lan\cdots\ran$ denotes event average. The second equality is due to the fact that the event average of $\c$ is zero. Thus in Cu + Au collisions, the measured directed flow is likely to be charge dependent. This may provide us a possibility to extract the electric conductivity of the produced matter by measuring the difference between the directed flows of positive and negative hadrons because $d_E$ is proportional to the electric conductivity $\s$ [See \eq{dE}]. This idea has been exploited in Ref.~\cite{Hirono:2012rt} (see also Ref.~\cite{Voronyuk:2014rna}) and we will not discuss it further.

Our main concern will be the two-particle correlator (\ref{gam}). The single-particle distribution (\ref{single_dis}) results in a two-particle distribution in the form~\footnote{The two-particle distribution may contain also irreducible component $C(q_1,\f_1;q_2,\f_2)$ which cannot be factorized into the product of single particle distributions. The mechanisms that underlie $C(q_1,\f_1;q_2,\f_2)$ include, for example, the transverse momentum conservation or local charge conservation.
We will discuss their effects in Sec.~\ref{concl}.}
\begin{eqnarray}
\label{two_dis}
f_2(q_1,\f_1;q_2,\f_2)=f_1(q_1,\f_1)f_1(q_2,\f_2),
\end{eqnarray}
from which one can derive
\begin{eqnarray}
\g_{q_1q_2}&=&\lan(v_1^0)^2\cos[2(\j_1-\j_\rp)]\ran\non&&
+(q_1+q_2)\lan d_E v_1^0\cos(\j_E+\j_1-2\j_\rp)\ran\non&&+ q_1 q_2\lan d_E^2\cos[2(\j_E-\j_\rp)]\ran\non&&+ q_1 q_2\lan d_B^2\cos[2(\j_B-\j_\rp)]\ran,
\end{eqnarray}
where the reaction plane angle $\j_\rp$ is defined as the angle between the impact parameter direction and the $x$ axis.
We see that the directed flow appears in the first two terms. For Au + Au collisions these two terms vanish because in that case the directed flow appears owing to fluctuation and does not correlate to the reaction plane. But for Cu + Au collisions, these two terms can be finite. However, if we study the difference between the opposite-sign and same-sign correlators, the directed flow terms are subtracted and we obtain
\begin{eqnarray}
\label{deltag}
\D\g&\equiv&\frac{1}{2}\lb2\g_{+-}-\g_{++}-\g_{--}\rb\non&=&-\frac{1}{2} \lan d_E^2\cos[2(\j_E-\j_\rp)]\ran-\frac{1}{2}\lan d_B^2\cos[2(\j_B-\j_\rp)]\ran.\non
\end{eqnarray}
Note that this correlator also subtracts other charge insensitive contributions (e.g.,the transverse momentum conservation) and has been extensively discussed in, for example, Refs.~\cite{Schlichting:2010qia,Bzdak:2012ia}.

Let us estimate the dipole strengths $d_E$ and $d_B$. Define $\bar{f}_1(q,\f)=f_1(q,\f)-f_1(q,\f)|_{\bE=\bB={\bf0}}$. Because the electric field is roughly in-plane as we have simulated (i.e. $\j_E\approx\j_\rp$), it transports charges across the $y-z$ plane during a proper time interval $0$ to $\t_E$ by an amount
\begin{eqnarray}
Q_{x>0}
&=&\int_0^{\t_E} d\t \t dy d\w \s E,
\end{eqnarray}
or
\begin{eqnarray}
\label{dqdeta1}
\frac{dQ_{x>0}}{d\w}
&=&\int_0^{\t_E} d\t \t dy \s E\non
&=&\frac{\t_E^2}{2}Y \s E
\end{eqnarray}
where $\t_E$ is the lifetime of the electric field, $\eta$ is the spacetime rapidity, $Y$ is a characteristic length scale of the overlapping region in $y$ direction, and $\s$ is electric conductivity. We have assumed that $\s$ and $E$ are constant in time and homogeneous in space.
Equation (\ref{dqdeta1}) should be equal to the integration of ${\bar f}_1(+,\f)-{\bar f}_1(-,\f)$ from $\j_\rp-\p/2$ to $\j_\rp+\p/2$, that is (To simplify the notation, we set $\j_\rp=0$),
\begin{eqnarray}
\label{dqdeta2}
\frac{dQ_{x>0}}{d\w}&=&\int_{-\p/2}^{\p/2} d\f\ls {\bar f}_1(+,\f)-{\bar f}_1(-,\f) \rs\non
&\approx&\frac{2N}{\p}d_E,
\end{eqnarray}
where $N=N_++N_-$ is the total multiplicity in a unit rapidity. Note that if there is a net charge, $\d N=N_+-N_-\neq 0$, the directed flow could also transport charges, that is why we use $\bar{f_1}$: it subtracts such non-EM-field induced charge transport. From \eq{dqdeta1} and \eq{dqdeta2}, we obtain
\begin{eqnarray}
\label{dE}
d_E&\approx &\frac{\p}{N}\frac{\t_E^2}{4}Y \s E.
\end{eqnarray}
In a similar way, we can estimate $d_B$ as
\begin{eqnarray}
\label{dB}
d_B&\approx &\frac{\p}{N}\frac{\t_B^2}{4}X \s_5 B,
\end{eqnarray}
with $\s_5$ the CME conductivity, $\t_B$ the lifetime of the magnetic fields, and $X$ a characteristic length scale of the overlapping region in $x$ direction. By adopting a simple hard-sphere model for the nucleus, we have
\begin{eqnarray}
\label{xyhardsphere}
X&\approx&R_{\rm Au}+R_{\rm Cu}-b,\non
Y&\approx&\frac{1}{b}\sqrt{[(R_{\rm Au}+R_{\rm Cu})^2-b^2][b^2-(R_{\rm Au}-R_{\rm Cu})^2]},
\end{eqnarray}
where $b$ is the impact parameter and $R_{\rm Cu}$ and $R_{\rm Au}$ are the radii of Cu and Au nuclei.

\subsection{Results}
\begin{figure}[!htb]
\begin{center}
\includegraphics[width=7cm]{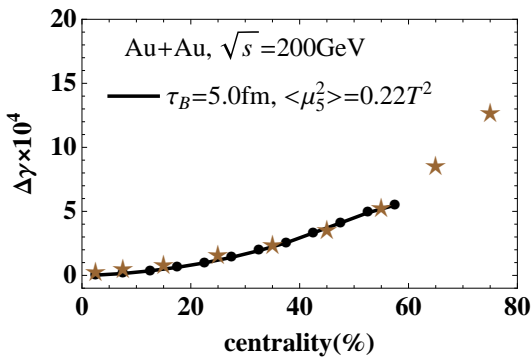}
\caption{(Color online) $\D\g$ as a function of centrality in Au + Au collisions. The star points are experimental results from ~\cite{Wang:2012qs}. The line is our fitting curve by using \eq{deltag}. We choose $T=350$ MeV and $\t_B=5$ fm in the fitting.}
\label{dg_AA}
\end{center}
\end{figure}

In the above modeling, the unknown parameters are the electric conductivity $\s$, the CME conductivity $\s_5$, and the lifetimes of fields $\t_{E,B}$.
For $\s$, we will use the recent lattice QCD results~\cite{Aarts:2007wj,Ding:2010ga,Amato:2013naa}
\begin{eqnarray}
\s=0.4C_{\rm EM}T,
\end{eqnarray}
with $C_{\rm EM}=\sum_fq_f^2=(2/3)e^2$ if we consider $u,d$ and $s$ quarks. We will take $T=350$ MeV in our numerical simulation. The CME conductivity is given by $\s_5=3C_{\rm EM}\m_5/(2\p^2)$. As it is $\s_5^2$ appearing in $\D\g$, thus instead of determining $\m_5$ itself we will determine $\lan\m_5^2\ran$ which measures the fluctuation of the chiral chemical potential because $\lan\m_5\ran=0$. For a given $\t_B$, we determine $\lan\m_5^2\ran$ (as a function of $T^2$) by fitting $\D\g$, \eq{deltag}, to the experimental result for Au + Au collisions reported by STAR collaboration~\cite{Wang:2012qs}. The results are shown in \fig{dg_AA} where we choose $\t_B=5$ fm~\footnote{The EM fields generated by the spectators decay very fast~\cite{Deng:2012pc}. In choosing $\t_B=5$ fm we have taken into account the fact that QGP is a good conductor. It is likely that the presence of a conducting medium may increase the lifetime of the magnetic fields owing to the Faraday induction, but there still remain debates, see discussions in Refs.~\cite{McLerran:2013hla,Tuchin:2013apa,Zakharov:2014dia}. In general, the lifetime of the electric fields is expected to be shorter than magnetic fields because the lack of an electric version of ``Faraday induction" to compensate the decay of the electric fields. In our numerical calculations, what is essential is the ratio $\t_E/\t_B$; thus once we fix $\t_B$ we treat $\t_E$ as a free parameter.}. At small centralities, the fitting is surprisingly good; at centrality larger than $60\%$ our model does not apply because the two nuclei do not overlap and Eqs.~(\ref{xyhardsphere}) (with $R_{\rm Cu}$ replaced by $R_{\rm Au}$) fails.
%%%%%%%%%%%%%%%%%%%%%%%%%%%%%%%%%%%%%%%%%%%%%%%%%%%%%%%%%%%%%%%%%%%%%%%
%%%%%%%%%%%%%%%%%%%%%%%%%%%%%%%%%%%%%%%%%%%%%%%%%%%%%%%%%%%%%%%%%%%%%%%

%%%%%%%%%%%%%%%%%%%%%%%%%%%%%%%%%%%%%%%%%%%%%%%%%%%%%%%%%%%%%%%%%%%%%%%
\begin{figure}[!htb]
\begin{center}
\includegraphics[width=7cm]{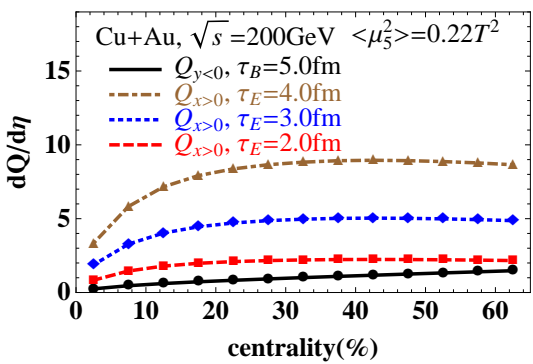}
\caption{(Color online) $dQ_{x>0}/d\eta$ and $dQ_{y<0}/d\eta$ versus centrality in Cu + Au collisions.}
\label{figdqdeta}
\end{center}
\end{figure}
%%%%%%%%%%%%%%%%%%%%%%%%%%%%%%%%%%%%%%%%%%%%%%%%%%%%%%%%%%%%%%%%%%%%%%%
%%%%%%%%%%%%%%%%%%%%%%%%%%%%%%%%%%%%%%%%%%%%%%%%%%%%%%%%%%%%%%%%%%%%%%%
\begin{figure}[!htb]
\begin{center}
\includegraphics[width=7cm]{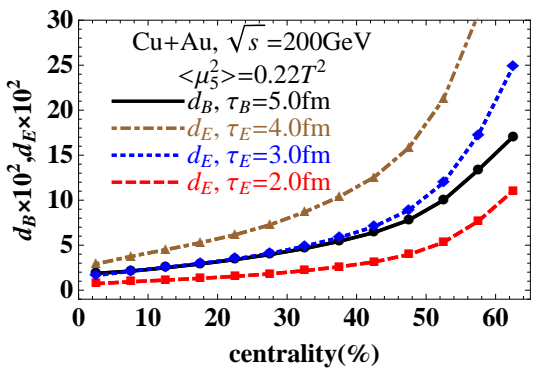}
\caption{(Color online) The in-plane dipole strength $d_E$ and the out-of-plane dipole strength $d_B$ versus centrality in Cu + Au collisions.}
\label{dedb}
\end{center}
\end{figure}
%%%%%%%%%%%%%%%%%%%%%%%%%%%%%%%%%%%%%%%%%%%%%%%%%%%%%%%%%%%%%%%%%%%%%%%
With $\lan\m_5^2\ran$ obtained, we apply \eq{deltag} to Cu + Au collisions and obtain $\D\g$ for Cu + Au collisions. Before we present the numerical results for $\D\g$, we in \fig{figdqdeta} and \fig{dedb} show $dQ_{x>0}/d\eta$, $dQ_{y<0}/d\eta$, and the dipole strengths $d_{E,B}$ as functions of centrality. As expected, as the lifetime of the electric fields increases more and more charges are transported across the $y-z$ plane. We note here that in plotting $dQ_{x>0}/d\eta$ in \fig{figdqdeta} we actually integrate the electric fields over the $y-z$ plane, thus $dQ_{x>0}/d\eta$ in \fig{figdqdeta} reflects the electric flux across $y-z$ plane. In contrast
to Ref.~\cite{Hirono:2012rt} but in agreement with Ref.~\cite{Voronyuk:2014rna}, we do not observe the sign-flipping feature of the electric field flux at any centrality. From \fig{dedb}, we observe that when the lifetime of the electric fields is longer than $3.0$ fm, the in-plane dipole induced by electric fields becomes stronger than the out-of-plane dipole induced by CME and we can expect some qualitative changes in $\D\g$ in this case.

\begin{figure}[!htb]
\begin{center}
\includegraphics[width=7cm]{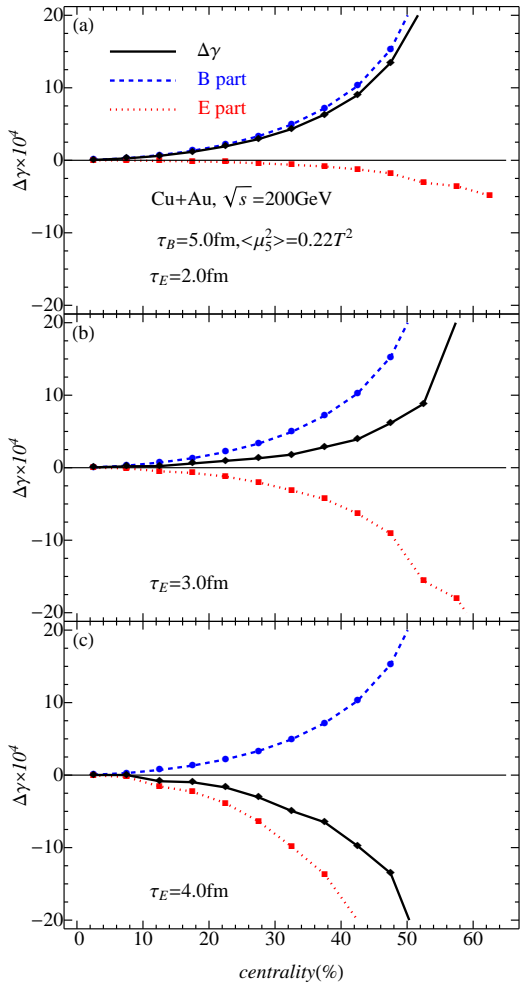}
\caption{(Color online) $\D\g$ versus centrality for Cu + Au collisions. The blue, red, and black curves represent the contributions from magnetic field, electric field, and the total, respectively.}
\label{dg_CA}
\end{center}
\end{figure}

The results for $\D\g$ are shown in \fig{dg_CA}. We see that, as expected, the in-plane electric fields give a negative contribution to $\D\g$ while the out-of-plane magnetic fields give a positive contribution (via the CME). For lifetime $\t_E=2$ fm (panel (a) of \fig{dg_CA}), we observe a suppression of $\D\g$ owing to the in-plane electric fields. This suppression becomes clearer for a longer lifetime $\t_E=3$ fm (panel (b) of \fig{dg_CA}). For an even longer lifetime, $\t_E=4$ fm, the electric fields dominate $\D\g$ and $\D\g$ becomes negative ( panel (c) of \fig{dg_CA}).
Based on these simulations, we hypothesize that the correlator $\D\g$ in Cu + Au may be much smaller or even reversed compared with that in Au + Au collisions.
%%%%%%%%%%%%%%%%%%%%%%%%%%%%%%%%%%%%%%%%%%%%%%%%%%%%%%%%%%%%%%%%%%%%%%%
%%%%%%%%%%%%%%%%%%%%%%%%%%%%%%%%%%%%%%%%%%%%%%%%%%%%%%%%%%%%%%%%%%%%%%%

\section{Summary and discussions}\label{concl}
In summary, by using HIJING model, we simulated the generation of the EM fields in Cu + Au collisions at $200$ GeV and studied the general properties of the EM fields, see Sec.~\ref{emfield}. We found that a strong electric field along the impact parameter direction exists in Cu + Au collisions. This in-plane electric field gives a negative contribution to the two-particle correlator $\D\g$ (defined in \eq{deltag}). By using a one-component model (i.e., by assuming $\D\g$ is driven by EM-field induced effects) we studied how the in-plane electric field influences $\D\g$ in Cu + Au collisions and found that it may strongly suppress or even change the sign of $\D\g$ compared with that in Au + Au collisions.

Some discussions are in order:\\
(1) The correlator $\D\g$ may receive other non-EM-field induced contributions (contained in the irreducible component, $C(q_1,\f_1;q_2,\f_2)$, of the two-particle distribution), potentially the LCC, which are not taken into account in the analysis in Sec.~\ref{seccorr}. However, inclusion of the non-EM-field induced contributions will not affect the role played by the in-plane electric field in Cu + Au collisions. The in-plane electric field always tends to drive an in-plane component of the charge dipole and thus tends to suppress $\D\g$.\\
(2) However, there is the possibility that $\D\g$ is dominated by non-EM-field induced effects and the CME and electric field play minor roles. In this case, one does not expect a strong suppression or reversion of $\D\g$ in Cu + Au collisions compared with that in Au + Au collisions. This provides us a feasible way to clarify how important the EM fields can be: If the experimental data in Cu + Au collisions really show a strong suppression or reversion of $\D\g$, it will strongly indicate that the EM fields play dominant roles in $\D\g$ (it will also indicate a large electric conductivity of the quark-gluon matter); otherwise it will be a hint for the $v_2$-driven interpretation of $\D\g$.\\
(3) The modeling of $\D\g$ in Sec.~\ref{seccorr} is illustrating, it can be improved by adopting a more realistic model, e.g., the two-component model~\cite{Bzdak:2012ia,Bloczynski:2013mca} for $\D\g$ (i.e., to include the contributions driven by $v_2$), by taking into account more realistic spacetime dependence of the EM fields, and by including the final state interactions~\cite{Ma:2011uma,Shou:2014zsa}. These will be our future tasks and the results will be presented in future publications.\\
(4) From \eq{deltag} we observe that the magnetic and the electric contributions to $\D\g$ are additive, thus it is interesting to construct suitable observables to separately characterize the in-plane and out-of-plane charge separations caused by the electric fields and magnetic fields, respectively. One possible observable is the charge dipole or multiple vector proposed in Ref.~\cite{Liao:2010nv,Bloczynski:2012en}, more detailed analysis will be future task.\\

\emph{Acknowledgments}---
We are grateful to J. Liao, G .L. Ma, A. Tang, and G. Wang for useful discussions.
XGH is supported by Shanghai Natural Science Foundation (Grant No. 14ZR1403000) and Fudan University (Grant No. EZH1512519). WTD is supported by the Independent Innovation Research Foundation of Huazhong University of Science and Technology (Grant No. 2014QN190) and the National Natural Science Foundation of China (Grant No. 11405066). We also acknowledge the support from the Key Laboratory of Quark and Lepton Physics (MOE), CCNU (Grant No. QLPL201417 and QLPL20122)

\end{document}